\documentclass[aip,jap,a4paper,reprint,amsmath,amssymb,floatfix]{revtex4-1}
\usepackage{amsmath}
\usepackage{bm}
\usepackage{graphicx}
\usepackage{siunitx}
\usepackage{ulem}
\usepackage{url}
\usepackage{latexsym,stmaryrd}
\usepackage[usenames,dvipsnames]{xcolor}

\begin{document}

\title{Influence of multi-electronic states on few-quantum-dot nanolasers}
\date{\today}
\author{J.~Liu}
\email{jin.liu@coer-scnu.org}
\affiliation{DTU Fotonik, Department of Photonics Engineering, Technical University of Denmark, Building 343, DK-2800 Kgs.~Lyngby, Denmark}
\affiliation{Niels Bohr Institute, University of Copenhagen, Blegdamsvej 17, DK-2100 Copenhagen, Denmark}
\affiliation{Present address: South China Academy of Advanced Optoelectronics, South China Normal University, 510006 Guangzhou, China}
\author{S.~Ates}
\affiliation{DTU Fotonik, Department of Photonics Engineering, Technical University of Denmark, Building 343, DK-2800 Kgs.~Lyngby, Denmark}
\affiliation{Present address: National Research Institute of Electronics and Cryptology, The Scientific and Technological Research Council of Turkey (TUBITAK), Gebze 41400, Turkey}
\author{M.~Lorke}
\affiliation{DTU Fotonik, Department of Photonics Engineering, Technical University of Denmark, Building 343, DK-2800 Kgs.~Lyngby, Denmark}
\affiliation{Present address: Bremen Center for Computational Materials Science, University of Bremen, Germany}
\author{J.~M{\o}rk}
\affiliation{DTU Fotonik, Department of Photonics Engineering, Technical University of Denmark, Building 343, DK-2800 Kgs.~Lyngby, Denmark}
\author{P.~Lodahl}
\author{S.~Stobbe}
\email{stobbe@nbi.ku.dk}
\affiliation{Niels Bohr Institute, University of Copenhagen, Blegdamsvej 17, DK-2100 Copenhagen, Denmark}


\begin{abstract}
We present an experimental and theoretical study on the gain mechanism in a photonic-crystal-cavity nanolaser with embedded quantum dots. From time-resolved measurements at low excitation power we find that four excitons are coupled to the cavity. At high excitation power we observe a smooth low-threshold transition from spontaneous emission to lasing. Before lasing emission sets in, however, the excitons are observed to saturate, and the gain required for lasing originates rather from multi-electronic transitions, which give rise to a broad emission background. We compare the experiment to a model of quantum-dot microcavity lasers and find that the number of emitters feeding the cavity must greatly exceed four, which confirms that the gain is provided by multi-electronic states. Our results are consistent with theoretical predictions.
\end{abstract}

\maketitle
Lasers can deliver coherent, narrow-band, and single-mode light and have become ubiquitous in contemporary technology. Nanometer-scale lasers may offer significant advantages because of large spectral separation of modes, compact sizes, and because the Purcell effect enables lowering the lasing threshold~\cite{Strauf2011}. Nanolasers may find applications in many areas including chip-to-chip optical communication~\cite{Kurosaka2010,Altug2006}, on-chip optical signal processing~\cite{Nozaki2012,Liu2010}, and biochemical sensing~\cite{Zhu2009,He2011}.

By virtue of the Purcell effect, the fraction of spontaneously emitted photons being emitted into the cavity mode, i.e., the $\beta$-factor, can be enhanced by either increasing the $Q$-factor or decreasing the mode volume of the cavity, which also increases the stimulated-emission rate~\cite{Gregersen2012}. Ultimately, the $\beta$-factor may approach unity~\cite{Lermer2013}, which implies that no threshold appears in the emitted intensity as a function of excitation power. Self-assembled quantum dots (QDs) embedded in photonic-crystal cavities provide a seemingly ideal setting to study nanolasing because of the simultaneous confinement of excitons and photons at the nanoscale~\cite{Noda2006} and lasing in this system was reported in Ref.~\onlinecite{Strauf2006}. However, a stark discrepancy between atomic laser models and experiment was also found: lasing was observed even when the excitons were completely detuned from the cavity. This non-resonant coupling has been the subject of intense experimental~\cite{Hennessy2007,Ates2009,Winger2009,Majumdar2012,Madsen2013} and theoretical~\cite{Naesby2008,Winger2009,Kaer2010} investigations and it is now understood~\cite{Strauf2011} that the dominant mechanisms are phonons~\cite{Madsen2013} and, at large excitation powers and/or large detunings, multi-electronic configurations due to hybridization with the wetting-layer states~\cite{Winger2009,Majumdar2012}, which may also be interpreted as Auger processes involving wetting-layer states.~\cite{Settnes2013}

Traditionally, laser oscillation is modelled by rate equations~\cite{Coldren_and_Corzine} where modified spontaneous and stimulated emission due to the cavity as well as the many-body effects of QDs are not included. It is only in recent years that a semiconductor model for QD-based microcavity lasers has been developed~\cite{Gies2007} and successfully applied to micropillar lasers to understand their coherence properties~\cite{Ates2008} and photon statistics~\cite{Ulrich2007}. However, a quantitative comparison between this model and photonic-crystal nanolasers at the few-QD level is still missing. In this work, we demonstrate few-QD nanolasing in photonic-crystal nanocavities and compare it to a semiconductor lasing model. The comparison of experiment and theory shows that lasing cannot be attributed to usual exciton transitions. Rather, it orginates from wetting-layer-mediated processes, which lead to a surprisingly high gain.

\begin{figure*}[t]
\centering
\includegraphics[width=\textwidth]{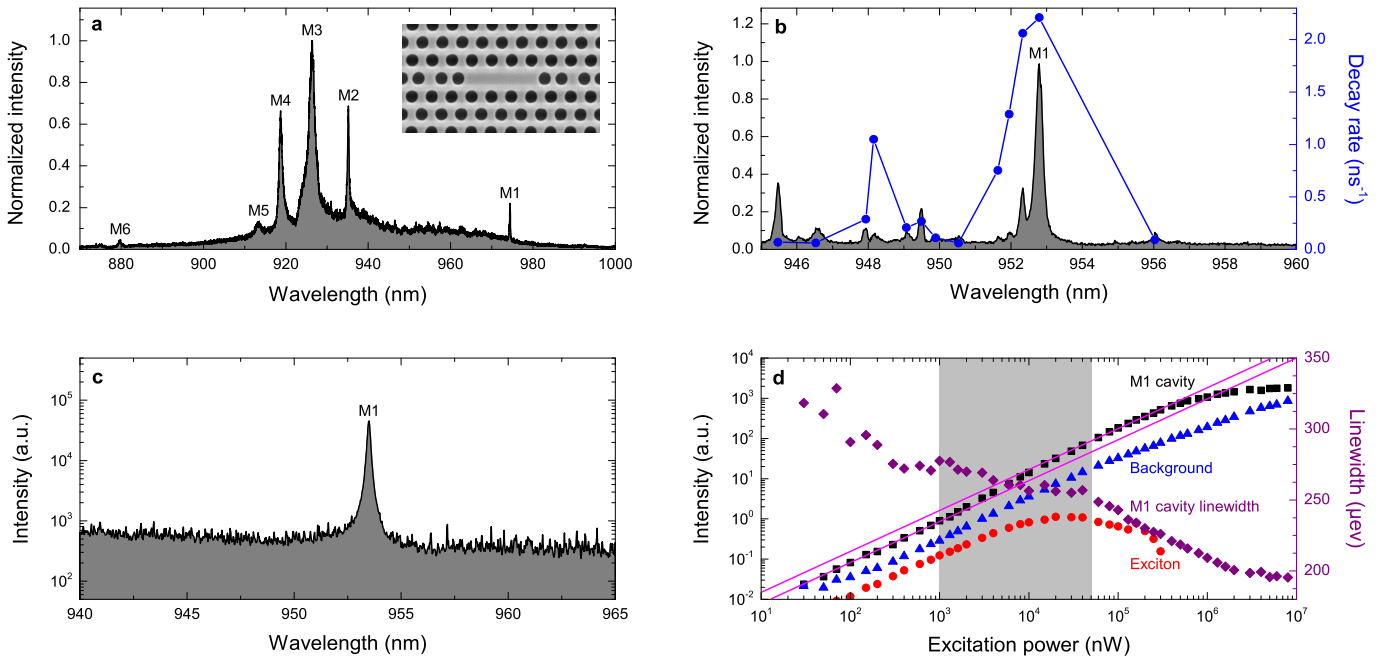}
\caption{Few-QD nanolasing. (a) Spectrum of an L3 cavity with embedded QDs at high-power (\SI{1}{\milli\watt}) above-band excitation with the characteristic cavity modes denoted M1-M6. The inset shows a scanning-electron micrograph of the sample.
(b)
Spectrum (black curve; left axis) around the M1-cavity obtained with M6-resonant excitation at \SI{100}{\nano\watt} showing the M1-cavity and a number of excitonic peaks whose decay rates (blue circles; right axis) have been measured using time-resolved spectroscopy.
(c)
Spectrum around M1 obtained with M6-resonant excitation at \SI{100}{\micro\watt} showing the M1-cavity peak on a pronounced background.
(d)
Integrated intensity as a function of excitation power for the M1-cavity mode (black squares), an exciton peak (red circles), and the background (blue triangles). The smooth s-shaped transition between linear regions (magenta lines) of the cavity power dependence is a characteristic signature of high-$\beta$ lasing. Evidently, the exciton saturates around threshold (gray region) and cannot provide the gain for the laser, which is provided by the background. The linewidth (purple diamonds; right axis) shows a small narrowing consistent with high-$\beta$ lasing.
}
\label{fig:1}
\end{figure*}

We have studied nanocavities obtained by leaving out three holes in a triangular-lattice photonic-crystal, i.e., L3-cavities~\cite{Akahane2003} as shown in the inset of Fig.~\ref{fig:1}(a). The membranes were \SI{154}{\nano\meter} thick and contained a single layer of embedded InAs QDs with a density of \SI{80}{\micro\meter^{-2}}. We characterized the samples by confocal diffraction-limited microphotoluminescence spectroscopy at a temperature of \SI{10}{\kelvin}. Figure~\ref{fig:1}(a) shows the emission spectrum of one cavity obtained with high-power above-band excitation; the six modes (denoted M1-M6) characteristic of the L3 cavity are efficiently excited by using the QD-ensemble as an internal broadband light source. Among all the cavity modes, the fundamental mode, M1, is the mode of our interest for laser oscillation since it has the highest Q-factor and the smallest mode volume.

In order to investigate how many exciton states are coupled to the M1-cavity mode, we applied a low excitation power of 100\:nW (measured before the microphotoluminescence objective) and recorded the spectrum around M1. Here and in the following we employ M6-cavity-resonant excitation in order to reduce undesired excitation of QDs outside the cavity. The decay rates of the exciton peaks were obtained by time-resolved spectroscopy. The result is shown in Fig.~\ref{fig:1}(b), where many single-exciton lines together with the M1-cavity mode are clearly observed. Four exciton lines are found to have significantly enhanced decay rates (faster than \SI{0.5}{\nano\second^{-1}}) due to the cavity as compared to the remaining excitonic states that are strongly inhibited by the band gap of the photonic crystal~\cite{Wang2011}. The power dependence of the four lines showed that they were due to single excitons.

The results shown in Fig.~\ref{fig:1}(b) seem to indicate that only four exciton states feed the cavity but this conclusion is incorrect because lasing sets in only at much higher excitation powers. For an excitation power of \SI{100}{\micro\watt}, the excitons are saturated and the spectrum is dominated by the cavity mode on a broad background as shown in Fig.~\ref{fig:1}(c). The background is due to multi-electronic states as discussed in Refs.~\onlinecite{Winger2009,Strauf2011}. We show below that at this excitation power the laser is above threshold.

The excitation-power dependence of the different spectral features is shown in Fig.~\ref{fig:1}(d). The M1-mode exhibits an s-shaped variation with excitation power bounded by linear regions and, for high excitation power, saturation, which is characteristic of high-$\beta$ lasing with a smooth threshold around the region marked in gray in Fig.~\ref{fig:1}(d). At high power levels the emission saturates. The excitons begin to saturate already in the region marked in gray, i.e., below the threshold, indicating that they are not (solely) providing the gain for lasing. The background integrated a spectral range corresponding to the linewidth of the cavity (measured off resonance), on the other hand, does not saturate, which could indicate that the gain is actually provided by the background, i.e., the multi-electronic states. This assertion is corroborated by our theoretical analysis described later. The cavity linewidth decreases with increasing excitation power but only a modest linewidth narrowing is observed because for high $\beta$, a significant fraction of the photons in the cavity are spontaneously emitted resulting in a reduced coherence as compared to macroscopic low-$\beta$ lasers~\cite{Ates2008,Strauf2011}.

We model our data by a microscopic QD-laser model introduced in Ref.~\onlinecite{Gies2007}. Within the cluster-expansion technique, the carrier-photon and photon-photon correlations are included and the light-matter interaction is treated at a microscopic level. This method has been used to study input-output characteristics as well as photon statistics in QD-based microcavity systems~\cite{Wiersig2009} allowing to extract the $\beta$-factor and the effective number of QD-excitons emitting into the lasing mode. In order to obtain a quantitative understanding of the experiment we have performed independent measurements of the most important parameters entering the model. Firstly, the $Q$-factor of the cavity was obtained ($Q=8500$) by fitting the cavity mode at the lasing threshold. Secondly, the light-matter coupling rate of the four coupled excitons was extracted from their decay rates. We note that while biexcitons can have a strong influence on input-ouput curves~\cite{Gies2011} they cannot explain the observed linewidth behavior.

\begin{figure}
\centering
\includegraphics[width=\columnwidth]{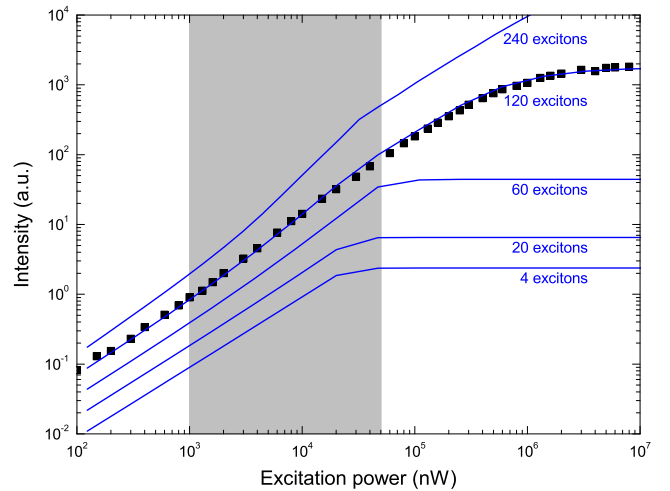}
\caption{Comparison of the input-output curves between the experiment (black squares) and the model (blue lines). Clearly, the model cannot fit the data for small numbers of emitters although only four excitons were found to be coupled to the cavity at low excitation power. By varying the number of emitters in the model as indicated in the plot we find that 120 emitters must be included to fit the experiment. The theoretical curves have been vertically offset for clarity.}
\label{fig:4}
\end{figure}

The input-output curve resulting from the procedure outlined above is shown in Fig.~\ref{fig:4}. Evidently, the gain provided by four excitons is insufficient to reach the lasing threshold. We therefore increased the number of excitons in the model and assumed a light-matter coupling strength equal to the average of the values obtained for the four excitons. We found a good agreement between the experiment and the model by using 120 excitons resulting in a $\beta$-factor of 0.4. This points to the fact that lasing in this device is indeed rather driven by the strong background emission observed directly in Fig.~\ref{fig:1}(c) than by actual intra-QD transitions and it is consistent with Fig.~\ref{fig:1}(d), where the exciton emission quenches whereas the integrated background emission shows a region of superlinear increase. The background is too spectrally wide to stem from phonon coupling and indicates rather transitions involving the continuum in the wetting layer and thus not excitons but rather multi-electronic states~\cite{Winger2009,Majumdar2012,Settnes2013}. We conclude that a gain equivalent to that of 120 excitons is required to model our data, which is much larger than the number of excitons coupling to the cavity at low excitation power. The actual number of electron-hole states involved in the lasing oscillation may be orders of magnitude larger but with correspondingly lower average light-matter coupling strengths.

In conclusion, we have studied lasing from nanocavities with few embedded QDs and compared the experimental data to a semiconductor model for QD-based microcavity lasers. The power dependence of the excitons, cavity, and background shows that only the background can provide the gain. This is confirmed by the quantitative analysis showing that a few excitons cannot provide enough gain for the lasing in such nanolasers and in fact the number of emitters needed is much higher than the number of excitons coupled to the cavity below threshold. Thus, our study confirms the picture in which the gain in few-QD nanolasers is provided by multi-electronic configurations where wetting-layer states play a major role.

We thank H.~Thyrrestrup and K.~H.~Madsen for fruitful discussions. We furthermore gratefully acknowledge the Danish Council for Independent Research (Natural Sciences and Technology and Production Sciences), the European Research Council (ERC consolidator grant), and the Villum Foundation (NATEC center of excellence) for financial support.

\end{document}